\documentclass[aps,prr,twocolumn,superscriptaddress,amsfonts,amssymb,amsmatht]{revtex4}\usepackage{xcolor}
\usepackage{ulem}
\usepackage[colorlinks=true,linkcolor=blue,citecolor=blue,urlcolor=blue]{hyperref}
\usepackage{tabularx}
\usepackage{multirow}
\usepackage{gensymb}
\usepackage{graphicx}
\usepackage{dcolumn}
\usepackage{bm}
\DeclareGraphicsExtensions{.ps,.eps,.pdf}
\DeclareGraphicsExtensions{.jpg,.png}

\begin{document}

\title{Bulk charge density wave and electron-phonon coupling in superconducting copper oxychlorides}

\author{L. Chaix}
\email[Corresponding author: ]{laura.chaix@neel.cnrs.fr}
\affiliation{Univ. Grenoble Alpes, CNRS, Grenoble INP, Institut Néel, 38000 Grenoble, France}

\author{B. Lebert}
\affiliation{Department of Physics, University of Toronto, Toronto, Ontario, M5S 1A7, Canada}

\author{H. Miao}
\affiliation{Condensed Matter Physics and Materials Science Department, Brookhaven National Laboratory, Upton, New York 11973, USA}
\affiliation{Materials Science and Technology Division, Oak Ridge National Laboratory, Oak Ridge, TN, USA.}

\author{A. Nicolaou}
\affiliation{Synchrotron SOLEIL, Saint-Aubin, BP 48, 91192 Gif-sur-Yvette, France}

\author{F. Yakhou}
\affiliation{European Synchrotron Radiation Facility, 71 Avenue des Martyrs, CS40220, F-38043 Grenoble Cedex 9, France}

\author{H. Cercellier}
\affiliation{Univ. Grenoble Alpes, CNRS, Grenoble INP, Institut Néel, 38000 Grenoble, France}

\author{S. Grenier}
\affiliation{Univ. Grenoble Alpes, CNRS, Grenoble INP, Institut Néel, 38000 Grenoble, France}

\author{N. B. Brookes}
\affiliation{European Synchrotron Radiation Facility (ESRF), B.P. 220, F-38043 Grenoble Cedex, France}

\author{A. Sulpice}
\affiliation{Univ. Grenoble Alpes, CNRS, Grenoble INP, Institut Néel, 38000 Grenoble, France}

\author{S. Tsutsui}
\affiliation{Japan Synchrotron Radiation Research Institute (JASRI), SPring-8, 1-1-1 Kouto, Sayo, Hyogo 679-5198, Japan}

\author{A. Bossak}
\affiliation{European Synchrotron Radiation Facility (ESRF), B.P. 220, F-38043 Grenoble Cedex, France}

\author{L. Paolasini}
\affiliation{European Synchrotron Radiation Facility (ESRF), B.P. 220, F-38043 Grenoble Cedex, France}

\author{D. Santos-Cottin}
\affiliation{Department of Physics, University of Fribourg, 1700 Fribourg, Switzerland}

\author{H. Yamamoto}
\affiliation{Laboratory for Materials and Structures, Tokyo Institute of Technology, 4259 Nagatsuta, Midori-ku, Yokohama, 226-8503, Japan}

\author{I. Yamada}
\affiliation{Department of Materials Science, Graduate School of Engineering, Osaka Prefecture University1-2 Gakuen-cho, Naka-ku, Sakai, Osaka 599-8570, Japan}

\author{M. Azuma}
\affiliation{Laboratory for Materials and Structures, Tokyo Institute of Technology, 4259 Nagatsuta, Midori-ku, Yokohama, 226-8503, Japan}
\affiliation{Kanagawa Institute of Industrial Science and Technology, Ebina 243-0435, Japan}

\author{T. Nishikubo}
\affiliation{Laboratory for Materials and Structures, Tokyo Institute of Technology, 4259 Nagatsuta, Midori-ku, Yokohama, 226-8503, Japan}

\author{T. Yamamoto}
\affiliation{Laboratory for Materials and Structures, Tokyo Institute of Technology, 4259 Nagatsuta, Midori-ku, Yokohama, 226-8503, Japan}

\author{M. Katsumata}
\affiliation{Laboratory for Materials and Structures, Tokyo Institute of Technology, 4259 Nagatsuta, Midori-ku, Yokohama, 226-8503, Japan}

\author{M. P. M. Dean}
\affiliation{Condensed Matter Physics and Materials Science Department, Brookhaven National Laboratory, Upton, New York 11973, USA}

\author{M. d’Astuto}
\email[Corresponding author: ]{matteo.dastuto@neel.cnrs.fr}
\affiliation{Univ. Grenoble Alpes, CNRS, Grenoble INP, Institut Néel, 38000 Grenoble, France}
\affiliation{Tokyo Tech World Research Hub Initiative (WRHI), Institute of Innovative Research, Tokyo Institute of Technology, 4259 Nagatsuta, Midori-ku, Yokohama, Kanagawa 226-8503, Japan}

\date{\today}

\begin{abstract}
Bulk charge density waves are now reported in nearly all high-temperature superconducting cuprates, with the noticeable exception of one particular family: the copper oxychlorides. Here, we used resonant inelastic X-ray scattering to reveal a bulk charge density waves in these materials. Combining resonant inelastic X-ray scattering with non-resonant inelastic X-ray scattering, we investigate the interplay between the lattice excitations and the charge density wave, and evidence the phonon anomalies of the Cu-O bond-stretching mode at the charge density wave wave-vector. We propose that such electron-phonon anomalies occur in the presence of dispersive charge excitations emanating from the charge density wave and interacting with the Cu-O bond-stretching phonon. Our results pave the way for future studies, combining both bulk and surface probes, to investigate the static and dynamical properties of the charge density wave in the copper oxychloride family.
\end{abstract}

\pacs{PACS }

\maketitle

\section{Introduction}

Charge density waves (CDWs) have been predicted in cuprates not long after the discovery of high-temperature superconductivity \cite{Zaanen1989, Machida1989} and are now considered as a generic property of the high-temperature superconducting (HTS) cuprates, although their microscopic origin and relation with the pseudogap and superconductivity remain unclear. In fact, experimental studies of CDWs in HTS cuprates were initiated with bulk-sensitive scattering measurements of La-based cuprates \cite{Tranquada1995} and surface-sensitive scanning tunnelling microscopy on Bi-based cuprates and copper oxychlorides (Ca$_{2-x}$Na$_{x}$CuO$_{2}$Cl$_{2}$) \cite{Hoffman2002, Hanaguri2004}. Then, the evolution and diversification of the techniques fertilized the experimental field of CDW orders in cuprates, providing deeper insights on a larger scale of compounds \cite{Shen2005, Wu2011, Thampy2014, Comin2016, Miao2019, Chaix2017, Li2020, Frano2020, Lee2021}. While all these studies indicate an ubiquitous CDW, they also point toward different characteristics which depend on the material. For instance, experimental observations have shown that, for different cuprate compounds, the CDW is short or long-range, coexists or not with an associated spin-order, or that its wave-vector increases or decreases with doping \cite{Comin2016, Frano2020}.  Such discrepancies have raised substantial questions about the nature of the CDW formation mechanism in HTS cuprates and stimulated scenarios involving either real-space local interactions or Fermi surface nesting \cite{Tranquada1995, Shen2005, Comin2014, Zaanen1989, Poilblanc1989}. Experimental studies have collected evidence for both scenarios and no consensus has been reached so far. 

\begin{figure}
\resizebox{8.6cm}{!}{\includegraphics{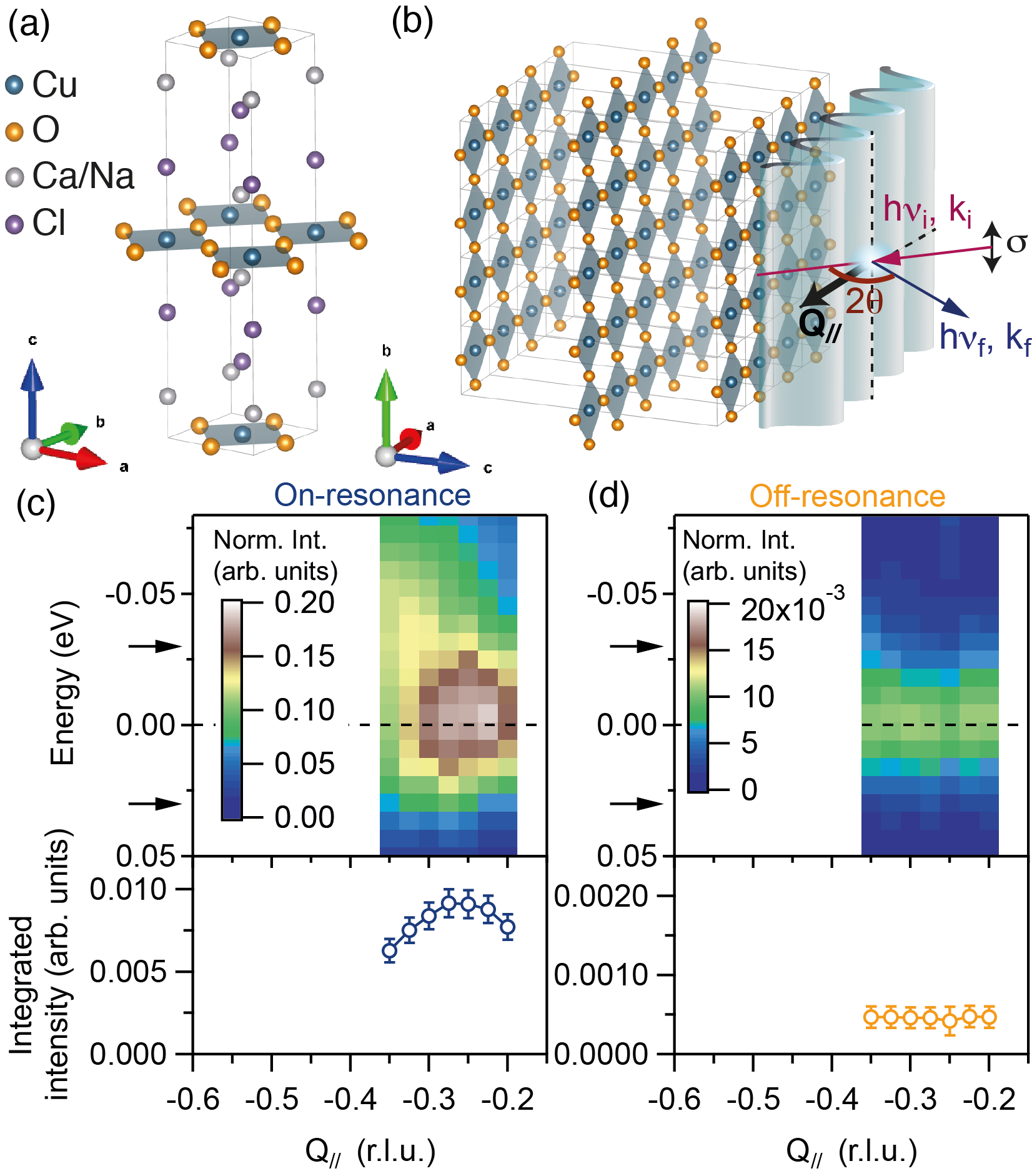}}
\caption{(a) Crystal structure of Na-CCOC \cite{Baptiste2018}. \textit{a}, \textit{b} and \textit{c} are the lattice axes of the sample. Cl ions are located on the apical sites. The four surrounding O form the square coordination of Cu, which is highlighted in blue. (b) Sketch of the RIXS scattering geometry with respect to the crystal structure of Na-CCOC and with a CDW. (c), (d) RIXS intensity maps recorded on- (c) and off- (d) resonance (top panels) as function of the energy and $Q_{\parallel}$. The integrated intensities in the quasi-elastic regions (indicated with black arrows on the RIXS intensity maps) are plotted on the bottom panels.}
\label{fig1}
\end{figure}

Combining surface and bulk probes on the same material can provide information on the CDW formation mechanism and its relation with both the superconductivity and the pseudogap. This approach, including, for example, scanning tunnelling microscopy, angle-resolved photoemission spectroscopy and resonant X-ray scattering, was applied to the Bi-based family. Such studies revealed a CDW which is consistent in surface and bulk and proposed possible connections to the fermiology \cite{Wise2008, Comin2014, Neto2014, Peng2018}. Ca$_{2-x}$Na$_{x}$CuO$_{2}$Cl$_{2}$ (Na-CCOC), whose surface electronic structure has been thoroughly explored \cite{Hanaguri2004, Shen2005}, is another promising system to combine surface and bulk-sensitive measurements to study the CDW, which exhibits unique characteristics. Interestingly this compound has a simple tetragonal $I4/mmm$ crystal structure [Fig. \ref{fig1} (a)] without any structural instabilities \cite{Vaknin1997} and presents strong 2D electronic properties \cite{Waku2004}, representing an ideal test bench to study the intrinsic physics from CuO$_{2}$ planes \cite{Shen2005}. Although this system was among the first HTS cuprates in which scanning tunnelling microscopy measurements successfully revealed a surface CDW, observed over a doping range from $x = 0.08$ to $x = 0.12$ \cite{Hanaguri2004, Fujita2014}, the existence of such order in bulk remains an open question. Indeed, a previous X-ray study, in which no bulk CDW was detected, suggested either its short-ranged nature or that the CDW is pinned at the surface \cite{Smadici2007, Brown2005}. Despite 25 years of research on bulk CDW in hole-doped cuprates, all converging toward its ubiquity and significance in the phase diagram, Na-CCOC remains an exception. This is an issue that deserves clarification with state-of-the-art X-ray scattering methods such as the highly-sensitive and energy-resolved soft X-ray resonant inelastic X-ray scattering (RIXS) technique.

In this study, we combine high energy-resolution RIXS at the Cu L$_{3}$-edge [Fig. \ref{fig1} (b)] and non-resonant inelastic X-ray scattering (IXS) measurements to investigate CDW and electron-phonon coupling anomalies in Na-CCOC with $x \sim 0.1$. Our results provide the first direct observation of a bulk and incommensurate CDW in this material. Concomitant to this observation, our RIXS and IXS data revealed anomalies on the Cu-O bond-stretching phonon characterized by (i) a softening and broadening at the CDW wave-vector, (ii) a RIXS intensity anomaly at $Q_{\text{A}} > Q_{\text{CDW}}$ and (iii) a funnel-like RIXS intensity emanating from the CDW. Together these findings, which are not observed in the undoped sample, point toward CDW-induced electron-phonon coupling anomalies that are interpreted in terms of dispersive charge excitations emanating from the CDW and interacting with the Cu-O bond-stretching phonon.

\section{Methods}

\subsection{Single-crystal synthesis}

Single-crystal synthesis of undoped copper oxychloride, $i.e.$ Ca$_{2}$CuO$_{2}$Cl$_{2}$ sample (CCOC), is described in Ref. \cite{Baptiste2018}. For the single-crystal synthesis of Na-doped copper oxychloride, $i.e.$ Ca$_{2-x}$Na$_{x}$CuO$_{2}$Cl$_{2}$ samples (Na-CCOC), we used the following precursors: CaCO$_{3}$ (99.99\%), CuO (99.99\%), CaCl$_{2}$ (99.99\%), NaClO$_{4}$ (99.9\%) and NaCl (99.99\%). First, we prepared a stoichiometric Ca$_{2}$CuO$_{2}$Cl$_{2}$ powder by a solid state reaction of CaCO$_{3}$, CuO, and CaCl$_{2}$ as described in previous works \cite{Hiroi1994, Kohsaka2002, Yamada2005}. In an argon filled dry box, we mixed the resulting Ca$_{2}$CuO$_{2}$Cl$_{2}$ powder with NaClO$_{4}$, NaCl and CuO precursors in a molar ratio of 1:0.2:0.2:0.2. We charged the mixture in cylindrical Pt capsules and then set the capsules in high-pressure cells. Since it was shown that the synthesis pressure is related to the Na content, we compressed the pressure cell between 3.5 to 4 GPa in a cubic anvil type high-pressure apparatus in order to get underdoped Na-CCOC single crystals. The capsules were heated up to 1250$\degree$C at a rate of 10$\degree$C/min, kept at this temperature for 1 h and then slowly cooled down to 1050$\degree$C at a rate of 10–20$\degree$C/h. After heat treatment, we released the pressures. We obtained single crystals with a Na doping concentration of $x \sim 0.1$ ($T_{\text{C}}$ $\sim 14$ K) whose crystal structure is shown on Fig. \ref{fig1} (a). The samples were characterized by X-ray diffraction and magnetic susceptibility measurements [see Appendix A, Fig. \ref{fig7} (a)].

\subsection{Resonant inelastic X-ray scattering  measurements at the Cu L$_{3}$-edge}

\begin{figure}[h]
\resizebox{8.6cm}{!}{\includegraphics{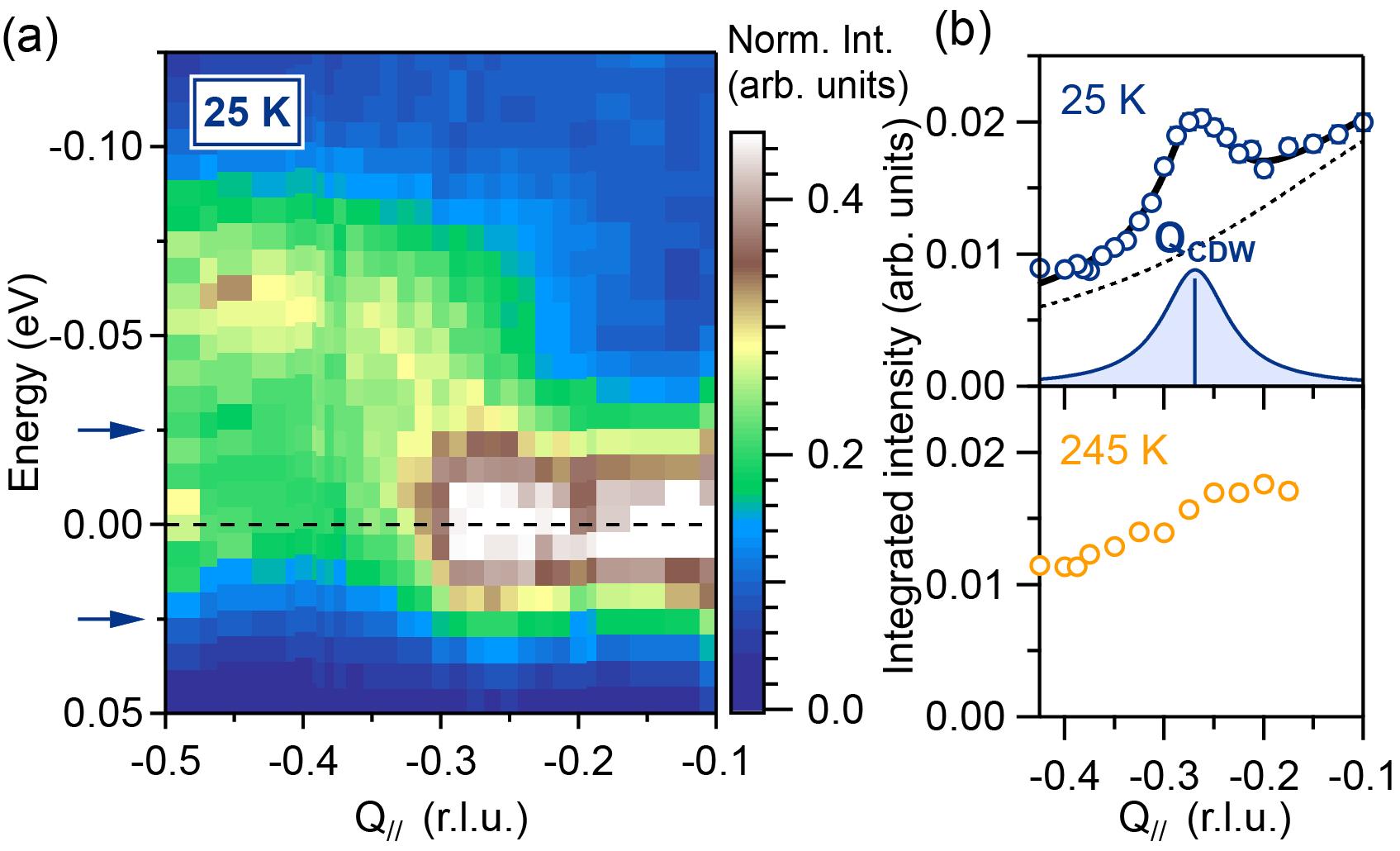}}
\caption{(a) RIXS intensity maps recorded at 25 K, as function of the energy and $Q_{\parallel}$. (b) Momentum-dependence of the integrated quasi-elastic intensity at 25 K (top) and 245 K (bottom). The quasi-elastic region is defined as between the two blue arrows on (a). The black solid line is a fit to the data consisting of two Lorentzians, one for the CDW (in blue) and the other for the tail of the specular reflection peak at $Q_{\parallel} = 0$ (dotted line), plus a constant. $Q_{\text{CDW}}$ is indicated with a vertical solid line.}
\label{fig2}
\end{figure}

Resonant inelastic X-ray scattering (RIXS) measurements were performed on the ERIXS spectrometer \cite{Brookes2018} at the ID32 beamline of the European Synchrotron Radiation Facility (ESRF). To avoid hygroscopic damage of the surface, the samples were cleaved under Ar atmosphere right before loading them into the high-vacuum chamber containing the four-circle diffractometer. All RIXS data shown in this report have been recorded with the incident photon energy tuned to the maximum of the Cu L$_{3}$-edge absorption curves, $i.e.$ at the resonance [see Appendix A, Fig. \ref{fig7} (b) for the corresponding absorption curve near the Cu L$_{3}$-edge], except data shown in Fig. \ref{fig1} (d), which were recorded off-resonance. The energy resolution was approximately $\Delta$E $\sim$ 50 meV for negative $Q_{\parallel}$ measurements of Na-CCOC, $\Delta$E $\sim$ 60 meV for positive $Q_{\parallel}$ measurements of Na-CCOC, except for the five first scans ($Q_{\parallel}$ = 0.15 to $Q_{\parallel}$ = 0.25 r.l.u.), where the energy resolution was $\Delta$E $\sim$ 80 meV, and $\Delta$E $\sim$ 65 meV for negative $Q_{\parallel}$ measurements of CCOC. The energy resolution was checked by recording RIXS spectra on silver paint at each $Q_{\parallel}$ position. The experimental geometry is sketched in Fig. \ref{fig1} (b). 

For all the RIXS data shown in this manuscript, we used a linear vertical x-ray polarization ($\sigma$-polarization), $i.e.$ perpendicular to the scattering plane. Na-CCOC sample was oriented with the $ac$-plane parallel to the scattering plane. In order to have the sample precisely aligned, the Na-CCOC orientation matrix was obtained by measuring the (0 0 2) Bragg peak at 932 eV and the (-1 0 1) Bragg peak at 1680 eV using the four-circle diffractometer inside the high-vacuum RIXS chamber. During the RIXS measurements, the scattering angle was fixed at 2$\theta$= 149.5$\degree$ giving a total momentum transfer $\vert Q \vert$ = 0.91 \AA$^{-1}$ with $Q = k_i – k_f$. Its component in the $ab$-plane, that is the projected momentum transfer $Q_{\parallel}$, was changed by rotating the sample around the $b$-axis, allowing to probe the dispersion of the excitations within the CuO$_2$ plane. Therefore, the projected momentum transfer $Q_{\parallel}$ was along the $<1 0 0>$ direction, varying from (0.1, 0) to (0.5, 0) in reciprocal lattice units (2$\pi$/a). All the RIXS data shown in this manuscript are plotted as function of the projected in-plane momentum transfer $Q_{\parallel}$ (along the \textit{a}-axis) in reciprocal lattice units (r.l.u.). 

We chose a convention where negative (positive) $Q_{\parallel}$ measurements correspond to grazing-incidence (emission) geometry. Note that we mainly measured RIXS spectra at negative $Q_{\parallel}$, $i.e.$ grazing-incidence geometry. Positive $Q_{\parallel}$ measurements, $i.e.$ grazing-emission geometry, were also performed on Na-CCOC at 25 K to confirm the results (see Appendix C). For the low-temperature data, RIXS spectra were recorded at 25 K, whereas, for the high-temperature measurements, RIXS spectra were measured at 245 K.

\subsection{Inelastic X-ray scattering measurements}

High-resolution inelastic X-ray scattering (IXS) was performed at the ID28 beamline of the ESRF \cite{Masciovecchio1996, Verbeni2008}, with complementary measurements performed at the BL35XU beamline of SPring-8 \cite{Baron2000}. Note that one data point at $Q$ = (3.065, 0.003, -0.073) has been recorded at the BL35XU beamline (SPring-8), whereas the data taken on the ID28 beamline (ESRF) concern the full dispersion along $Q$ = (3 0 0) - ($q$ 0 0). 9 in-line analyzers were installed in the IXS spectrometer of ID28, while BL35XU uses a 3 x 4 analyzer. The Si(9 9 9) backscattering setup was chosen: the incident X-ray energy was 17.793 keV and the energy resolution was 3 meV on average, with small variations ($<$ 10\%) depending on the analyzer. In particular we found 2.96(7) meV for analyzer 2 of ID28, from the fit of the elastic line width of our own data. To reduce the tails of the low energy modes and avoid multiple scattering, the IXS spectra were measured at 25 K using a closed cycle refrigerator. The resolution ($\sim$ 3 meV) is decoupled from the incident energy, $i.e.$ depends from the Darwin width of the main monochromator and is constant over all the exchanged energy range. This guarantee that the broadening of the optic phonon is effective. Background observed between 60 and 100 meV mostly comes from elastic scattering from the sample. Grease was used to protect the hygroscopic samples from air and to mount them on the copper sample holders in the cryostat. 

\begin{figure*}
\resizebox{17.2cm}{!}{\includegraphics{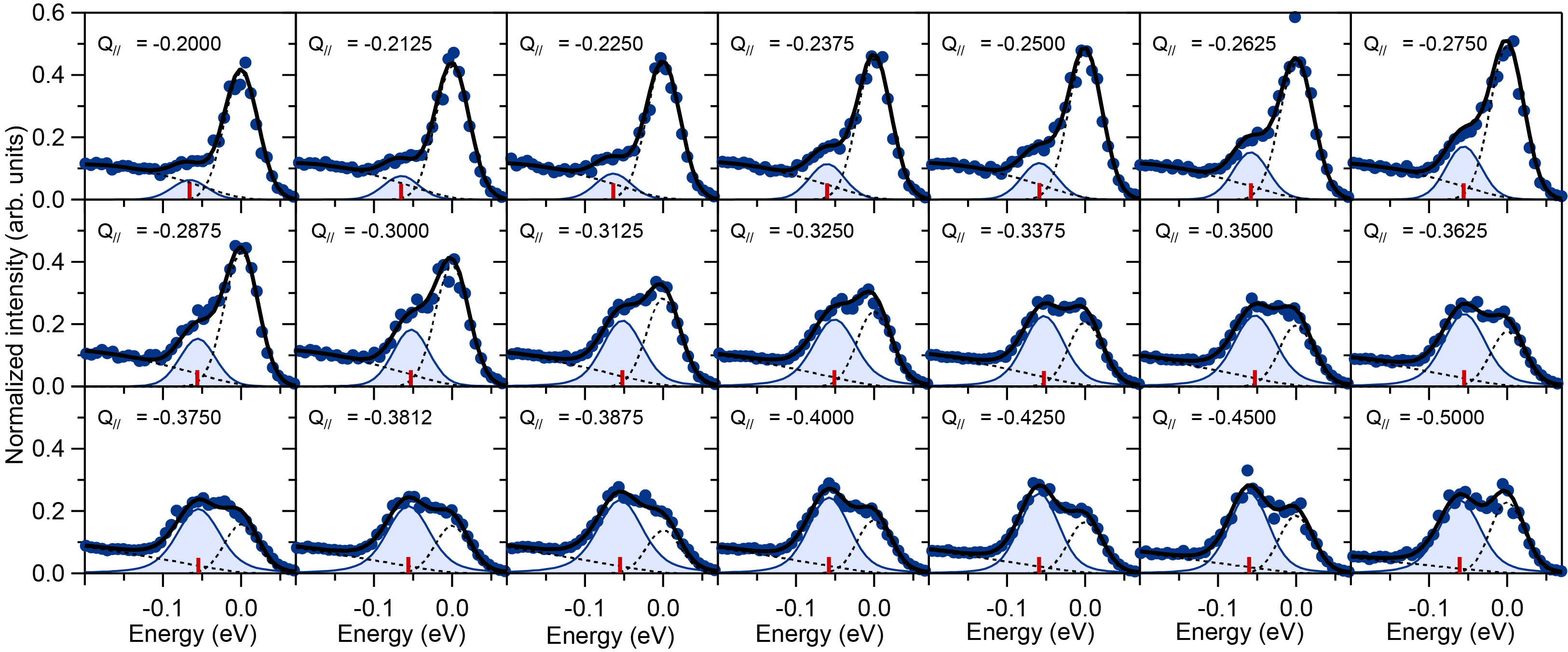}}
\caption{Fits of the RIXS spectra measured at 25 K in Na-CCOC. Energy-loss spectra plotted over the low-energy region demonstrating the quality of the fits. The model used takes into account a Gaussian for the elastic peak, a Lorentzian for the Cu-O bond-stretching phonon and an anti-symmetrized Lorentzian for the magnetic excitations, convoluted with the Gaussian energy resolution function ($\Delta$E $\sim$ 50 meV). The data are shown with markers and the resulting fits are displayed with black solid lines. The fitted Cu-O bond-stretching phonons are highlighted with filled areas. The red vertical lines indicate the fitted phonon position.}
\label{fig3}
\end{figure*}

\section{Results}

\subsection{Bulk charge density wave in Na-CCOC}

The RIXS technique has recently emerged as decisive to detect incipient electronic orders such as CDW and its corresponding excitations/fluctuations \cite{Comin2016, Miao2019, Chaix2017, Li2020, Lin2020, Lee2021, Miao2017, Arpaia2019, Yu2020}. It is suitable to detect weak signals. This is due to the combination of the resonant process, which increases the sensitivity to the valence electrons, and the high energy-resolution now available, which can discriminate low-energy excitations from quasi-elastic contributions. Figure \ref{fig1} (c) displays a RIXS intensity map recorded on-resonance and at 25 K in Na-CCOC. Enhanced quasi-elastic intensity peaking at an in-plane momentum transfer of $Q_{\parallel} \sim -0.26$ r.l.u. (reciprocal lattice units) is detected. In contrast, the intensity drops by two orders of magnitude in the same energy- and momentum-range, when the incident photon energy is off-resonance [Fig. \ref{fig1} (d)]. The observed quasi-elastic peak is therefore dominated by the resonant process, indicating that it is associated with the electronically active Cu states.

More quantitative assessments can be achieved by analyzing the high-resolution ($\Delta E \sim 50$ meV) and high statistics RIXS intensity map displayed in Fig. \ref{fig2} (a). The RIXS intensity integrated over a small energy window (-25 meV $<$ E $<$ 25 meV) around the zero-energy loss is shown in Fig. \ref{fig2} (b). A quasi-elastic peak is confirmed. The peak position is found at $Q_{\text{CDW}} \sim -0.27 \pm 0.01$ r.l.u., while its FWHM $\sim  0.08 \pm 0.01$ r.l.u., indicates a correlation length of $\xi$ $\sim 15$ $-$ $20$ $\text{\AA}$. Note that measurements at positive $Q_{\parallel}$ confirm these observations (see Appendix C). This peak is the signature of an incommensurate short-range CDW, that is static or quasi-static in nature (on a timescale of $\sim$ 160 fs, limited by the experimental energy resolution). Upon increasing the temperature to 245 K, the CDW peak is absent, as shown in Fig. \ref{fig2} (b), and only the specular scattering background remains, indicating that the CDW correlation increases when decreasing the temperature.

\begin{figure*}
\includegraphics[width=\textwidth]{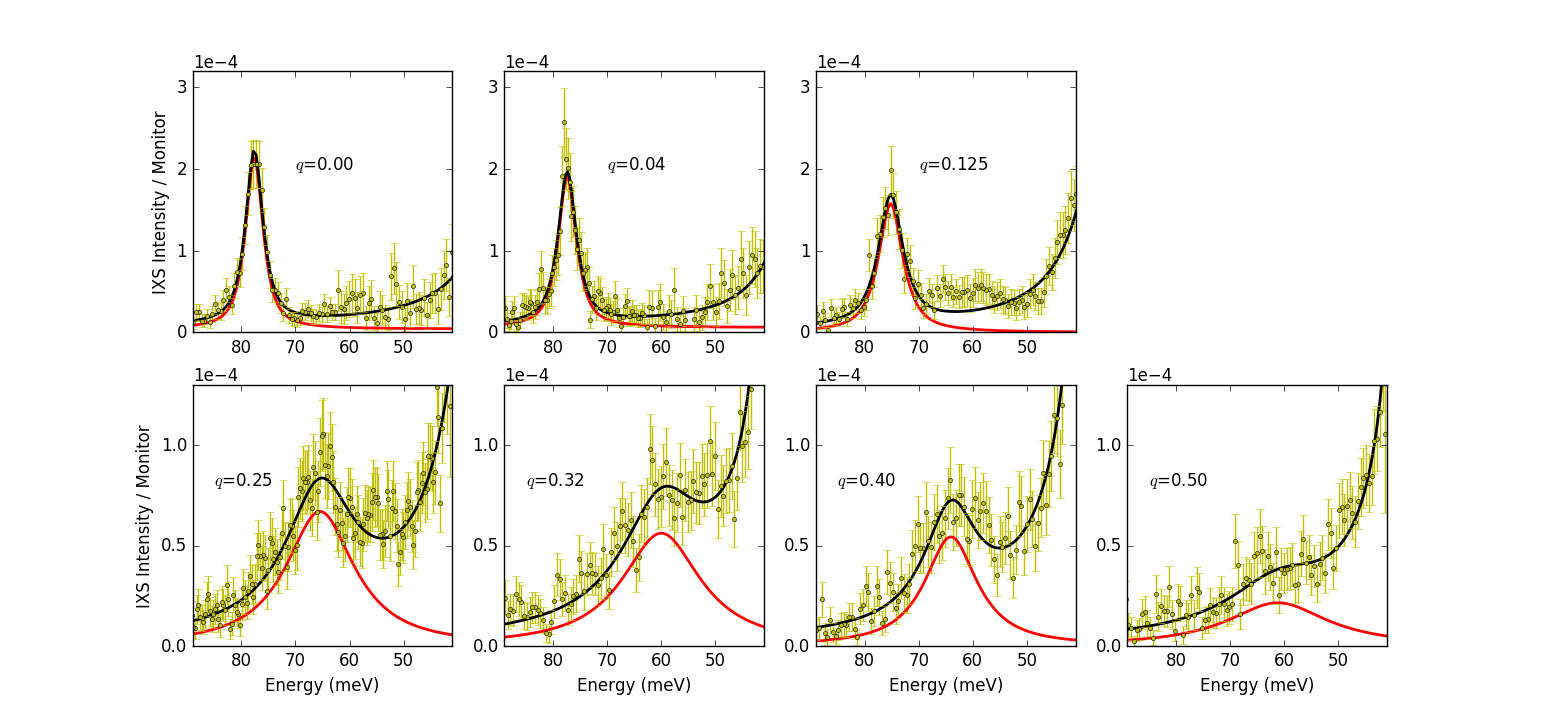}
\caption{Fits of the IXS spectra measured at 25 K at the ID28 beamline of ESRF in Na-CCOC plotted over the Cu-O bond-stretching phonon energy region demonstrating the quality of the fits. Pseudo-Voigt line shapes were used to fit the BS phonon mode. The data are shown with markers and the resulting fits are displayed with black solid lines. The fitted BS phonons are highlighted with red lines. Normalized Intensity corresponds to the ratio of IXS Intensity with the beam monitor. }
\label{fig4}
\end{figure*}

\subsection{Cu-O bond-stretching phonon and electron-phonon coupling anomalies} 

We now focus on the inelastic features and investigate the interplay between the lattice excitations and the CDW. Indeed, the study of the phonon spectra can provide insight on the driving force behind its formation mechanism, possibly revealing Kohn anomaly resulting from Fermi surface nesting or any momentum-dependent electron-phonon coupling \cite{Zhu2015}. Inelastic neutron scattering and IXS have identified anomalies in the phonon spectra close to the CDW wave-vectors in several HTS cuprates, suggesting an intimate relation with the lattice \cite{Uchiyama2004, Reznik2006, Graf2008, Astuto2008, Reznik2010, Astuto2013, Tacon2014, He2018, Miao2018}. These anomalies could be connected to the origin of the CDW, and have been discussed in term of CDW-induced modification of the lattice \cite{Lin2020}, hybridization between different phonon modes \cite{Graf2008, He2018} or even the presence of collective charge excitations \cite{Park2014}. Despite of being intensively studied over the past 30 years, the very nature of these phonon anomalies remains puzzling in HTS cuprates, calling for a new paradigm in phonon investigation. Because of its sensitivity to electron-phonon coupling and the recent technical progress made on its energy-resolution, the RIXS is now becoming a key complementary technique to investigate phonon spectra and electron-phonon coupling in HTS cuprates. 

\begin{figure*}
\resizebox{16cm}{!}{\includegraphics{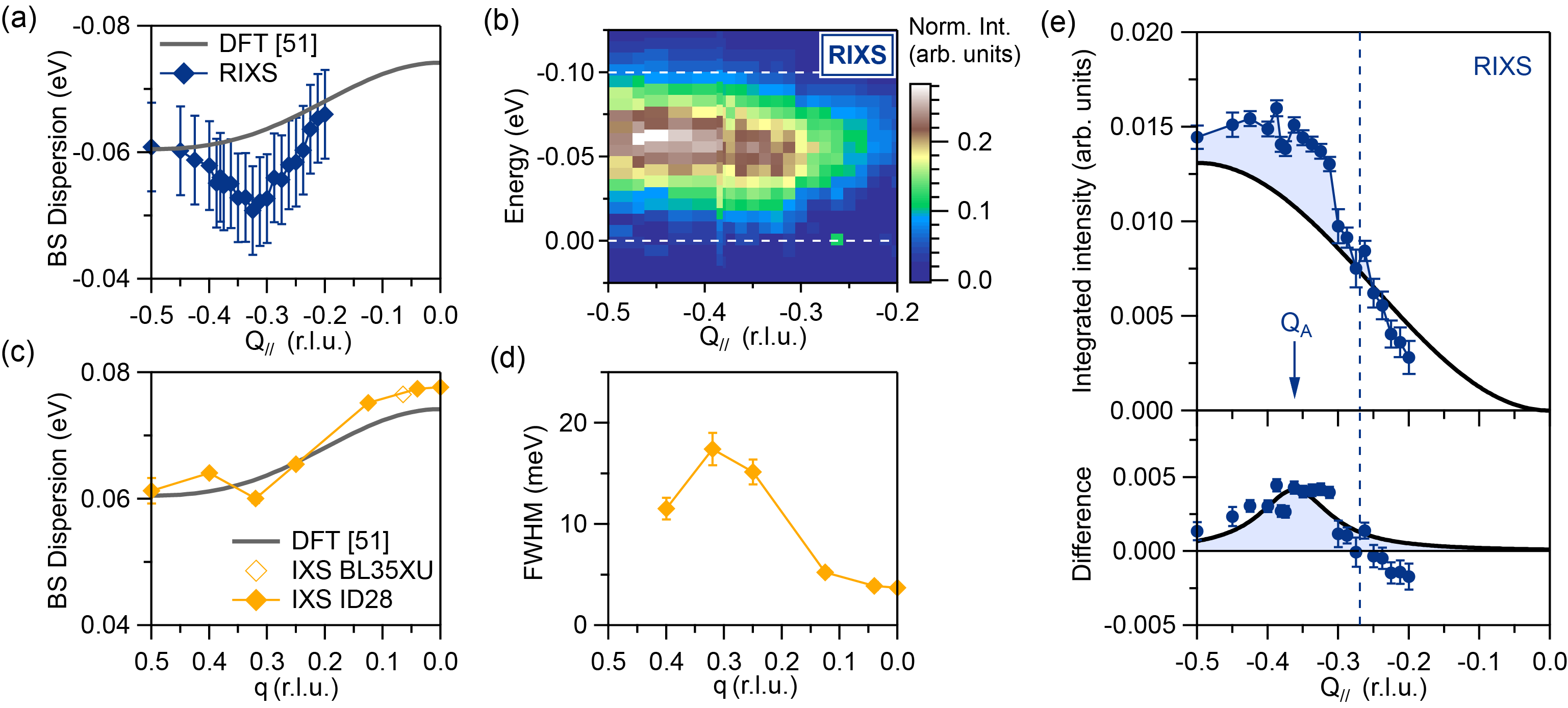}}
\caption{(a) RIXS BS phonon dispersion at 25 K. density functional theory calculations, extracted from ref. \cite{Lebert2020}, are shown with a grey solid line. (b) RIXS intensity map at 25 K, with subtracted elastic and paramagnon peaks. The raw data are the same than those shown in Fig. \ref{fig2} (a). (c), (d) IXS BS phonon dispersion (c) and width (d) at 25 K. IXS spectra have been measured at both BL35XU (SPring-8) and ID28 (ESRF) beamlines (see Methods). Note that in contrast to the peak position, the fitted FWHM at $q$ = (0.5 0 0) was a parameter not stable in the fit (with very large error bars), which strongly depends on the model used (either pseudo-Voigt or damped harmonic-oscillator). (e) Momentum-dependence of the RIXS integrated intensity of the elastic/paramagnon-subtracted data at 25 K with $\text{sin}^{2}(\pi Q_{\parallel})$ curve (black solid line). The integrated energy window corresponds to the white dashed lines in (b). The bottom panel represents the difference between the integrated intensity and the $\text{sin}^{2}(\pi Q_{\parallel})$ curve, with Lorentzian fit (solid line). $Q_{\text{CDW}}$ and $Q_{\text{A}}$ are indicated with a dashed line and arrow respectively.}
\label{fig5}
\end{figure*}

As presented in Fig. \ref{fig3}, the energy-loss RIXS spectra at 25 K show an excitation centered at $\sim$ 60 meV, between the elastic peak and the tails of the magnetic excitations observed at higher energy. The dispersion of this excitation is shown on Fig. \ref{fig5} (a) and was obtained by plotting the peak position (see red ticks on Fig. \ref{fig3}) as function of $Q_{\parallel}$. To identify this excitation, we have performed IXS measurements on another Na-CCOC sample from the same batch. As shown in Fig. \ref{fig4} displaying IXS spectra at 25 K, a peak corresponding to a phonon mode is observed at 60 meV, whose dispersion is plotted in Fig. \ref{fig5} (c). Its energy and dispersion agree well with the 60 meV excitation observed in the RIXS data, indicating that both excitations, $i.e.$ the one seen in RIXS and the one seen in IXS, have common origin. A comparison to density functional theory calculations shown on Fig. \ref{fig5} (a) and (c) and other HTS cuprates indicate that this mode corresponds to the Cu-O BS phonon \cite{Uchiyama2004, Reznik2006, Graf2008, Astuto2008, Reznik2010, Astuto2013, Pintschovius1991, McQueeney1999}. Interestingly, as seen by both RIXS and IXS, the BS phonon dispersion softens. This softening presents two components. First, a doping-induced softening in a form of a cosine-like dispersion along ($q$ 0 0), which has been already reported in a earlier IXS study \cite{Astuto2013} and reproduced well by the density functional theory. Then, a pronounced dip is observed at $Q \sim 0.3$ r.l.u., $i.e.$ close to the CDW wave-vector, which is not captured by the density functional theory \cite{Reznik2008, Lebert2020}. This dip coincides with a peak width broadening occurring between 0.2 and 0.4 r.l.u. [Fig. \ref{fig5} (d)] and resembles the phonon anomaly observed in conventional metallic systems such as NbSe$_2$, where the electron-phonon coupling is known to have a substantial implication on the formation of the CDW \cite{Zhu2015}. Note that the $Q_{\parallel}$ of the RIXS dispersion minimum is not exactly coinciding with $Q_{\text{CDW}}$, as also observed in La$_{2-x}$Sr$_x$CuO$_4$ \cite{Lin2020}. This could indicate that the phonon softening and CDW wave-vector relationship is more complex than expected. This could also originate from the $Q_{\parallel}$-dependent electron-phonon coupling which is higher at larger $Q_{\parallel}$ and tends to displace the minimum of the BS phonon dispersion \cite{Lin2020}. Note also that a previous IXS study on Bi2201 revealed a second phonon mode in the 40-80 meV energy range, whose dispersion crosses or anticrosses the BS phonon at $q$ = (0.25 0 0) \cite{Graf2008}. However, our IXS data do not indicate the presence of another mode in this energy range, which should be well separated to the BS phonon for spectra away from the crossing point (for instance for those close to the Brillouin zone center) and should display a significant dispersion [see Figure \ref{fig9}]. Our results suggest that the CDW affects the lattice dynamics, and in the present case the BS mode, by renormalizing its frequency and linewidth. This points toward an intimate connection between the electron-phonon coupling and the CDW that deserves further examination.

To study the role of the electron-phonon coupling, we can examine the RIXS phonon intensity. Indeed, this technique has recently been used to quantify the electron-phonon coupling strength $\lambda$ in several HTS cuprates, confirming a momentum dependent electron-phonon coupling that is strongest at the zone boundary for the BS mode \cite{Rossi2019, Braicovich2020}. This is due to the intrinsic nature of the RIXS process, in which the phonons are excited through the electron-phonon coupling interaction \cite{Ament2011, Devereaux2016}. Consequently, the RIXS phonon intensity is proportional to the electron-phonon coupling strength and can thereby reflect interactions between phonons and underlying charge excitations \cite{Chaix2017, Li2020, Lee2021}. As a function of the momentum, the electron-phonon coupling should scale as I$(Q) \propto \text{sin}^{2} (\pi \text{H}) + \text{sin}^{2} (\pi \text{K})$ for the BS mode \cite{Devereaux2016}, $i.e.$ $\text{sin}^{2} (\pi Q_{\parallel})$ for our scattering geometry. However, Fig. \ref{fig5} (e) shows that a deviation is observed in our data. Indeed, the momentum-dependence of the BS phonon integrated intensity presents an anomaly at $\vert Q_{\text{A}} \vert > \vert Q_{\text{CDW}} \vert$, as indicated by the difference plot shown in the bottom panel of Fig. \ref{fig5} (e), and confirmed by the analysis of the positive $Q_{\parallel}$ data (see Appendix C). A Lorentzian fit of the difference gives $Q_{\text{A}} \sim -0.36 \pm 0.02$ r.l.u. This deviation confirms the presence of electron-phonon coupling anomalies in the vicinity of the CDW. 

Similar electron-phonon coupling anomalies have been observed in La$_{1.8-x}$Eu$_{0.2}$Sr$_{x}$CuO$_{4+\delta}$ \cite{Peng2020, Wang2021} and Bi-based cuprates \cite{Chaix2017, Li2020, Lee2021} using RIXS. The former seems to occur in a narrow momentum region near the CDW wave-vector and has been discussed in term of enhanced electron-phonon coupling within the CDW phase that further stabilizes the CDW modulation \cite{Peng2020}. The latter however occurs for $\vert Q \vert > \vert Q_{\text{CDW}} \vert$ and has been attributed to originate from an interference between the BS phonon and charge excitations emanating from the CDW \cite{Chaix2017, Li2020, Lee2021}. Due to their dispersive nature, these excitations broaden in momentum with increasing energy to form a funnel-like intensity emanating from the CDW. Once crossing the BS phonon, an interference occurs that leads to an intensity anomaly in the RIXS data. Since this interference is dictated by the electron-phonon coupling, which increases monotonically with the momentum for the BS mode, the intensity anomaly is more visible for $\vert Q \vert > \vert Q_{\text{CDW}} \vert$, $i.e.$ where the electron-phonon coupling is strong \cite{Chaix2017, Li2020, Lee2021}. Our observation of an intensity anomaly occurring at $\vert  Q_{\text{A}} \vert > \vert Q_{\text{CDW}} \vert$ therefore favors the second scenario, $i.e.$ the interference between the BS phonon and charge excitations dispersing from the CDW. This is further supported by the RIXS intensity map presented in Fig. \ref{fig2} (a) showing an additional scattering intensity smoothly connecting the CDW and the BS phonon and forming a funnel-like intensity emanating from the CDW [Fig. \ref{fig5} (b)]. Together, these observations point toward the existence of dispersive excitations associated with the CDW. By connecting $Q_{\text{CDW}}$ and $Q_{\text{A}}$, we determined a velocity for these excitations: $v_{\text{CDW}} \sim 0.4$ eV$\cdot$\AA, consistent with the velocities found in Bi-based cuprates \cite{Chaix2017, Li2020}. This result differs from a recent RIXS study on La$_{2-x}$Sr$_x$CuO$_4$, in which the phonon softening has been attributed to a CDW-induced modification of the lattice without any coupling to electronic excitations \cite{Lin2020}. This might be an indication of the different electronic character of the CDW in La$_{2-x}$Sr$_x$CuO$_4$ relative to Na-CCOC.

\begin{figure}
\resizebox{8.6cm}{!}{\includegraphics{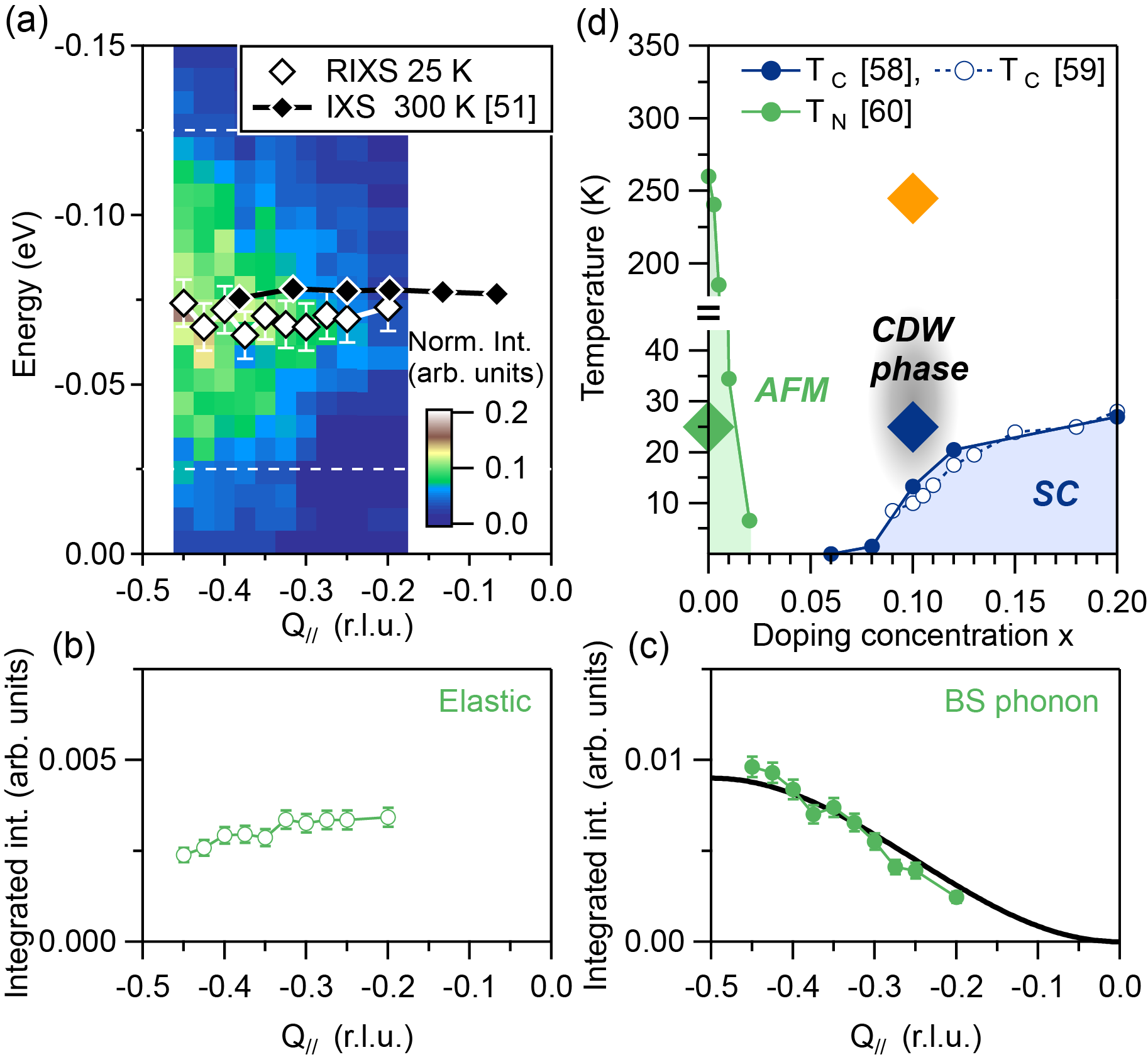}}
\caption{(a) Elastic and magnons-subtracted RIXS intensity map of CCOC at 25 K. The RIXS BS phonon dispersion is displayed with white diamonds. The black diamonds show the IXS BS phonon dispersion, from ref. \cite{Lebert2020}. (b) Momentum-dependence of the integrated quasi-elastic RIXS intensity of the CCOC data. The integrated energy window corresponds to the blue arrows on Fig. \ref{fig2} (a). (c) Momentum-dependence of the integrated RIXS intensity of the elastic and magnon-subtracted data of CCOC with $\text{sin}^{2} (\pi Q_{\parallel})$ curve (black solid line). The integrated energy window corresponds to the white dashed lines in (a). (d) Temperature-doping phase diagram of Na-CCOC. The onset temperatures, i.e. $T_N$ (anti-ferromagnetic, AFM) and $T_c$ (superconducting, SC), are from ref. \cite{Azuma2003, Zhigadlo2007, Ohishi2005}. The different doping concentrations and temperatures probed in our study are reported with filled diamonds.}
\label{fig6}
\end{figure}

The CDW nature of the observed anomalies is confirmed by recording RIXS spectra on the undoped Ca$_2$CuO$_2$Cl$_2$ (CCOC) sample. Fig. \ref{fig6} (a) presents the energy-momentum intensity map, with the fitted elastic peak and magnons subtracted. As expected, a good agreement is obtained by comparing the dispersion of the 70 meV RIXS excitation (white diamonds) to the one of the BS phonon measured by IXS on another CCOC sample from the same batch (black diamonds) \cite{Lebert2020}. The CCOC BS phonon is almost dispersionless, with no softening in the probed momentum range. In addition, the momentum-dependence of the quasi-elastic peak confirms the absence of the CDW in CCOC [Fig. \ref{fig6} (b)]. As anticipated, the phonon integrated intensity shown in Fig. \ref{fig6} (c) increases monotonically with $Q_{\parallel}$. More importantly, its momentum-dependence is well captured by the $\text{sin}^{2} (\pi Q_{\parallel})$ trend, consistent with predictions due to electron-phonon coupling for the BS phonon \cite{Devereaux2016, Braicovich2020}. These observation confirm that the phonon and electron-phonon coupling anomalies observed in Fig. \ref{fig5} exist only in doped CCOC, where the CDW is present.

\section{Discussion}

To visualize our results, the temperature-hole doping phase diagram of Na-CCOC is displayed in Fig. \ref{fig6} (d). Sodium substitution for calcium in CCOC dopes the Mott insulating state,inducing a metallic state with no long-range magnetic order and superconductivity at higher doping \cite{Azuma2003, Zhigadlo2007, Ohishi2005}. Our study unambiguously reveals a bulk CDW at 25 K in the underdoped region of the phase diagram, $i.e.$ close to the doping concentration of $x \sim 0.1$. Our results suggest that the CDW is short-ranged ($\sim 15 - 20$ \AA), comparable to the correlation lengths found in Bi-based cuprates, known for exhibiting chemical disorder and inhomogeneity \cite{Frano2020}. They also indicate that the CDW is incommensurate with $Q_{\text{CDW}} \sim -0.27 \pm 0.01$ r.l.u.

Our estimation of $Q_{\text{CDW}}$ from RIXS in Na-CCOC is slightly different than the one claimed in scanning tunnelling microscopy studies, discussing a commensurate CDW ($Q_{\text{CDW}}$ = 0.25), which is observed over a doping range from $x = 0.08$ to $x = 0.12$ \cite{Hanaguri2004}. In the following, we discuss several scenarios that could explain this discrepancy. First, it is important to note that the scanning tunnelling microscopy CDW wave-vector, estimated from the Fourier transform of the conductance maps, was given without error bars, contrary to our case. We cannot judge their error, but a closer inspection of their data indicates that, while the overall scanning tunnelling microscopy peaks seem centered on commensurate positions for the (2$\pi$, 0) directions, it is not the case for the (0, 2$\pi$) directions, where the maximums are peaked on incommensurate positions, close to the CDW wave-vector found in our RIXS study \cite{Hanaguri2004}. It is therefore possible that the scanning tunnelling microscopy data are not precise enough to distinguish if the surface CDW is precisely commensurate or incommensurate as found in our case. Another possibility is that the scanning tunnelling microscopy CDW is commensurate and that the difference between RIXS and scanning tunnelling microscopy lies in the difference between surface and bulk properties. In this scenario, the commensurate CDW would not be representative of the bulk electronic properties and would only be located at the surface, possibly stabilized by a soft surface phonon \cite{Brown2005}. While we cannot exclude the possibility of a material-specific feature here, such discrepancy was however not observed in Bi-based compounds, where combined X-ray and scanning tunnelling microscopy studies showed consistency between surface and bulk properties \cite{Comin2014, Neto2014}. An alternative scenario would be that the CDW is locally commensurate, locked to a $4a_{0}$ periodicity as probed by scanning tunnelling microscopy \cite{Hanaguri2004}, but appears incommensurate in X-ray studies, through discommensuration effects \cite{Mesaros2016}. This last proposal comes from recent analysis of the scanning tunnelling microscopy data in Bi-based cuprates where they revealed a doping independent local commensurate CDW with a $4a_{0}$ periodicity while scattering measurements identified an incommensurate global modulation with a wave-vector decreasing with increasing doping \cite{Mesaros2016, Webb2019, Zhao2019}. Similar recent observations have been made on the high-magnetic phase of YBCO using nuclear magnetic resonance \cite{Vinograd2021}. This scenario, which may be consistent with the doping-independent commensurate CDW wave-vector seen in Na-CCOC with scanning tunnelling microscopy \cite{Hanaguri2004}, requires however more studies to be confirmed. We therefore believe that our results will motivate further in depth investigations of the CDW in Na-CCOC combining both bulk and surface probes, as it has been done in Bi-based compounds \cite{Comin2014, Neto2014, Mesaros2016, Webb2019, Zhao2019}.

Concerning the inelastic features, our IXS measurements revealed anomalies in the BS phonon, resulting in a peak width broadening and a dip in its dispersion at $Q_{\text{CDW}}$, which is not reproduced by the density functional theory. We also evidenced that such lattice anomalies are concomitant to RIXS intensity anomalies which are characterized by a momentum-dependent electron-phonon coupling that deviates from the predictions for the Cu-O bond stretching mode and a funnel-like intensity emanating from the CDW. Together, these results point toward the existence of additional charge excitations dispersing from the short-range CDW at 25 K and interacting with the Cu-O bond stretching phonon at 60 meV. This extends the observation of such excitations outside the Bi-based family \cite{Chaix2017, Li2020, Lee2021}, whose nature has been discussed in term of a collective mode corresponding to SU(2) fluctuations \cite{Morice2018} or a continuum of quantum fluctuations that could melt the CDW below $T_{C}$ \cite{Lee2021}. This last scenario might open a new route for the CDW and superconductivity relationship, beyond the simple competing picture. It also points toward the presence of quantum critical points, which could be responsible for many unconventional properties found in these systems. By demonstrating that CDW excitations also exist in Na-CCOC, we offer another playground to study this scenario.

\section{Conclusion}

To summarize our main finding, the study we described in the present work : 
\begin{itemize}

\item it unambiguously reveals a bulk CDW at 25 K and close to the doping concentration of $x \sim 0.1$, which is short-ranged ($\sim 15 - 20$ \AA),and  incommensurate with $Q_{\text{CDW}} \sim -0.27 \pm 0.01$ r.l.u.;

\item it shows that the phonon anomalies seen by IXS, directly probing the $\mathcal{S}(\mathbf{q},\omega)$, are concomitant to RIXS intensity anomalies showing that electron-phonon coupling originates them; 

\item  as previously observed only in the Bi-based family \cite{Chaix2017, Li2020, Lee2021}, a funnel-like inelastic intensity emanate from the CDW. 

\end{itemize}

Together, these results point toward the existence of additional charge excitations dispersing from the short-range CDW at 25 K and interacting with the Cu-O bond stretching phonon at 60 meV

\section*{Acknowledgments}

RIXS experiment was performed at the ID32 beamline of the European Synchrotron Radiation Facility (ESRF, Grenoble, France) supported by a beamtime allocation under the proposal HC-3884. IXS measurements were performed at the ID28 beamline of ESRF supported by a beamtime allocation under the proposal HC-4310. Additional IXS measurements were conducted at BL35XU of SPring-8 (Japan) under the SPring-8 Budding Researchers Support Program (proposal No. 2019B1779). We thank O. Leynaud for access to the X’Press diffractometer platform as well as L.-M. Chamoreau and B. Baptiste for access to the “Plateforme Diffraction” at IPCM, Paris, and providing help for the single-crystal X-ray diffraction experiments. We thank L. Del-Rey and D. Dufeu for conceiving the sample holders and J. Debray for his help for the preparation of the RIXS experiment. We thank L. Laversenne and N. Bendiab for access to the glove-boxes. We thank P. Rodi\`ere and M.-H. Julien for fruitful discussions. We acknowledge financial support from ANR-DFG grant ANR-18-CE92-0014-03 ”Aperiodic”, IRS IDEX/UGA and Grants-in-Aid for Scientific Research 18H05208, and 19H05625, from the Japan Society for the Promotion of Science (JSPS).

\newpage

\section{Appendix}

\subsection{Sample preparations and characterizations}

\begin{figure}
\resizebox{8.6cm}{!}{\includegraphics{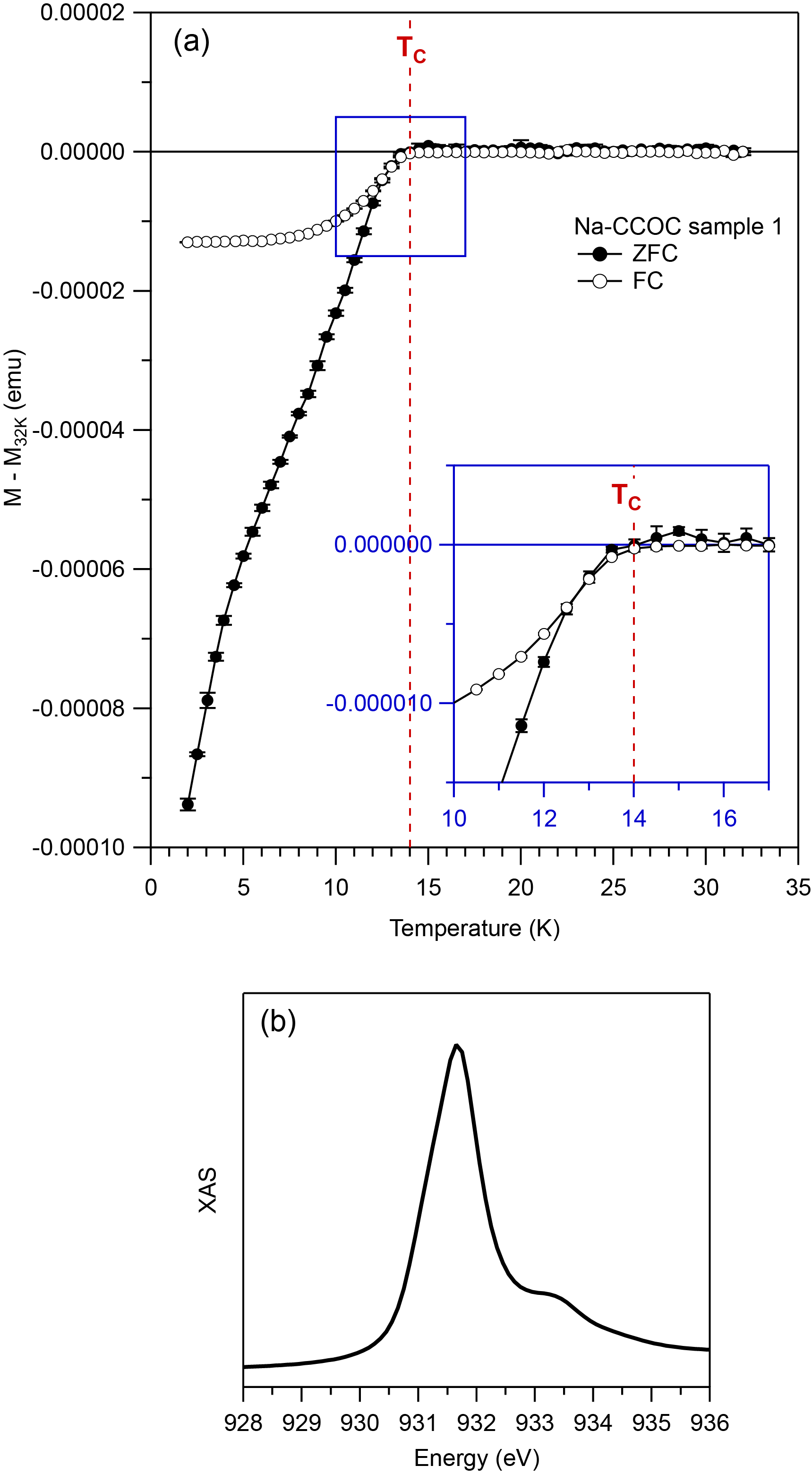}}
\caption{(a) Zero-field-cooling (ZFC) and field-cooling (FC) magnetization measurements vs temperature with a magnetic field of 46.3 Oe of the Na-CCOC sample 1 (sample measured with RIXS) across the superconducting transition. The $T_{C}$ corresponds to the onset of the diamagnetic signal. Insert shows an enlarged view of the signal within the blue box near $T_{C}$. (b) X-ray absorption spectrum (XAS) near the Cu L$_{3}$-edge of Na-CCOC.}
\label{fig7}
\end{figure}

\begin{figure*}
\resizebox{17.2cm}{!}{\includegraphics{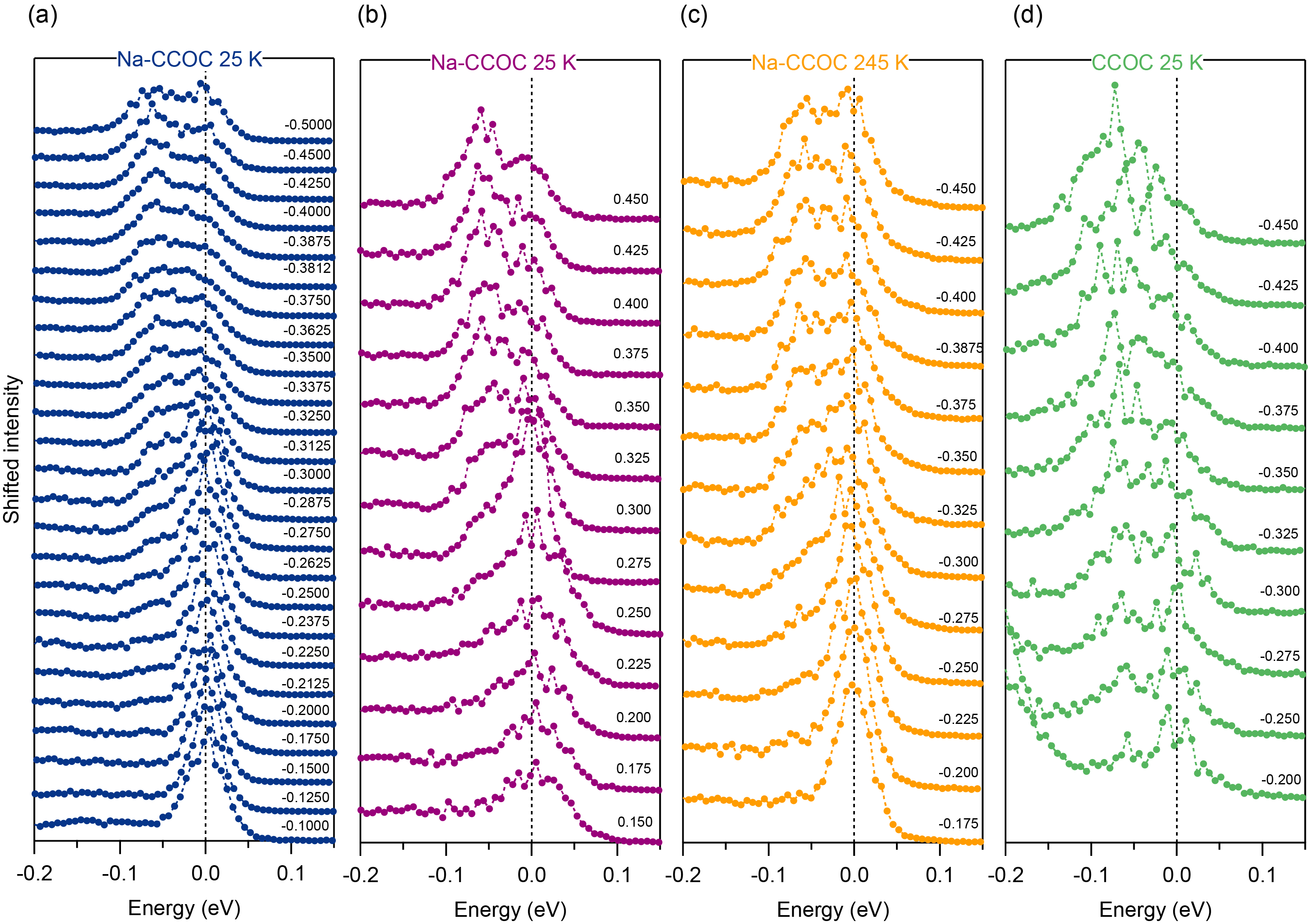}}
\caption{RIXS spectra of Na-CCOC sample 1 recorded at 25 K at negative $Q_{\parallel}$ with an energy resolution of $\Delta$E $\sim$ 50 meV (a) and positive $Q_{\parallel}$ with an energy resolution of $\Delta$E $\sim$ 60 meV (except for the five first scans, $i.e.$ $Q_{\parallel}$ = 0.15 to $Q_{\parallel}$ = 0.25 r.l.u., where the energy resolution was $\Delta$E $\sim$ 80 meV) (b). (c) RIXS spectra of Na-CCOC recorded at 245 K at negative $Q_{\parallel}$ with an energy resolution of $\Delta$E $\sim$ 50 meV. (d) RIXS spectra of CCOC recorded at 25 K at negative $Q_{\parallel}$ with an energy resolution of $\Delta$E $\sim$ 65 meV. The spectra are shifted vertically for clarity.}
\label{fig8}
\end{figure*}

For this study, we choose Ca$_{2-x}$Na$_{x}$CuO$_{2}$Cl$_{2}$ (Na-CCOC) samples with a doping concentration of $x \sim 0.1$ [$T_{\text{C}}$ $\sim 14$ K, see Fig. \ref{fig7} (a)], similarly to the previous scanning tunnelling microscopy studies in which a surface CDW was successfully reported. We used four different samples of Ca$_{2-x}$Na$_{x}$CuO$_{2}$Cl$_{2}$: three doped samples called Na-CCOC samples 1, 2 and 3, as well as one undoped one called CCOC sample. The three doped ones (Na-CCOC samples 1, 2 and 3) were from the same batch. The Na-CCOC sample 1 and CCOC sample were used to perform the RIXS measurements whereas the Na-CCOC sample 2 and 3 were used for the IXS measurements at ID28 beamline of ESRF and BL35XU beamline of SPring-8 respectively. The samples were pre-aligned before the measurements using single-crystal X-ray diffraction. Ca$_{2-x}$Na$_{x}$CuO$_{2}$Cl$_{2}$ is sensitive to humidity, so to avoid hygroscopic damages of the surface, they were encapsulated within two Kapton foils during the X-ray diffraction measurements. For the RIXS measurements, the samples were glued with silver paint on the sample-holders. Their manipulations and preparations were done inside glove-boxes.

\subsection{RIXS and IXS data analysis}

\begin{figure}
\resizebox{8cm}{!}{\includegraphics{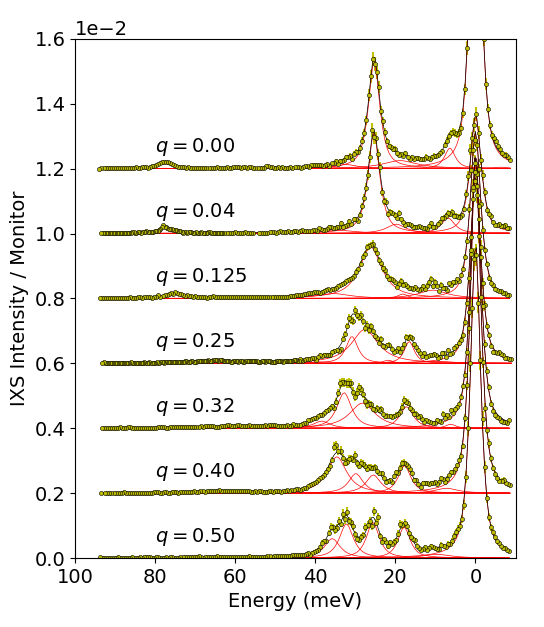}}
\caption{IXS spectra measured for different $Q=(3-q,0,0)$ and at 25 K in Na-CCOC sample 2. The data have been recorded on the ID28 beamline of ESRF. The intensity for each wave-vector below q=0.5 are shifted of 0.2x10$^{-2}$, so that the Brillouin zone center is at the top, and the zone boundary at the bottom. Fig. \ref{fig4} in the main text show a zoom of the same data, focussing on the details around the Cu-O bond-stretching mode energy range.}
\label{fig9}
\end{figure}

\begin{table}[!h]
\begin{center}
    \begin{tabular}{|c||c||c|}
    \hline      
    & & \\ 
       & Negative $Q_{\parallel}$ data & Positive $Q_{\parallel}$ data \\
    & & \\ 
    \hline  
    & & \\ 
  CDW & $Q_{\text{CDW}} \sim$ & $Q_{\text{CDW}}^{\dag} \sim$ \\
  wave-vector (r.l.u.) & $-0.27 \pm 0.01$ & $0.28 \pm 0.01$ \\
    & & \\
  CDW & FWHM $\sim$ & FWHM$^{\dag}$ $\sim$ \\
  width (r.l.u.) & $0.08 \pm 0.01$ & $0.06 \pm 0.01$ \\
    & & \\
  Correlation & $\xi$ $\sim 15$ & $\xi^{\dag}$ $\sim  20$ \\
  length (\AA) & &  \\
    & & \\
  Phonon & $Q_{\text{A}} \sim$ & $Q_{\text{A}}^{\dag}$ $\sim$ \\
  anomaly (r.l.u.) & $-0.36 \pm 0.02$ & $0.34 \pm 0.02$ \\
    & & \\ 
    \hline  
    \end{tabular}
\end{center}
\caption{Comparison of the wave-vectors, widths and correlation lengths of the CDW, as well as the wave-vectors of the phonon intensity anomalies from the analysis of the RIXS data recorded at negative and positive $Q_{\parallel}$.}
\label{tab1}
\end{table}

Fig. \ref{fig8} summarizes the RIXS spectra of our study, recorded on Na-CCOC at 25 K and 245 K at negative $Q_{\parallel}$ [Fig. \ref{fig8} (a) and (c)] and positive $Q_{\parallel}$ [Fig. \ref{fig8} (b)] and on CCOC at 25 K [Fig. \ref{fig8} (d)]. RIXS spectra shown in this report have been extracted with a Single Photon Counting algorithm using the RixsToolBox software \cite{Kummer2017}, except for Fig. \ref{fig1} (c) and \ref{fig1} (d), which display spectra extracted with traditional algorithm. The data have been normalized to I$_0$ and corrected to self-absorption effects using the formalism described in Ref. \cite{Minola2015, Rossi2019}, except for Fig. \ref{fig1} (c) and \ref{fig1} (d), where only I$_0$ normalization has been applied. The zero-energy loss position was precisely determined using the procedure described in Ref. \cite{Chaix2017}: each spectrum taken on the samples has been roughly align by a comparison to a spectrum acquired on silver paint near the samples, then the zero-energy position was finely adjusted through the fit of the elastic peak. The model used to fit each RIXS spectrum involves four components, convoluted with the Gaussian energy resolution function: a Gaussian function for the elastic peaks, a Lorentzian function for the Cu-O bond stretching phonons, an anti-symmetrized Lorentzian function for the magnetic excitations and a Gaussian background to account for the tail of the excitations at higher energy. Fig. \ref{fig3} demonstrates the quality of the fits. Note that the plotted fits are already convoluted with the energy resolution function. Similar fitting procedures and analysis are commonly found in the literature \cite{Chaix2017, Li2020, Peng2015}. Error bars for the RIXS phonon dispersion are estimated by the uncertainty in determining the zero energy loss [Fig. \ref{fig5} (a)]. Those for the RIXS integrated intensity plots are determined by Poisson statistics.

The IXS spectra recorded on ID28 and BL35XU were normalized to  to their respective beamline monitor. Fig. \ref{fig9} summarizes the corresponding IXS data. The spectra have been analyzed using a similar fitting procedure than the one described in \cite{Astuto2007}. Pseudo-Voigt line shapes were used to fit the Cu-O bond stretching phonon mode. We also checked the results using a damped-harmonic-oscillator model \cite{Fak1997} for the Cu-O bond stretching phonon mode, which overlap well within error-bars with the pseudo-Voigt model. Fig. \ref{fig4} demonstrates the quality of the fits. 
\\

\subsection{Complementary positive $Q_{\parallel}$ measurements}

\begin{figure*}
\resizebox{17.2cm}{!}{\includegraphics{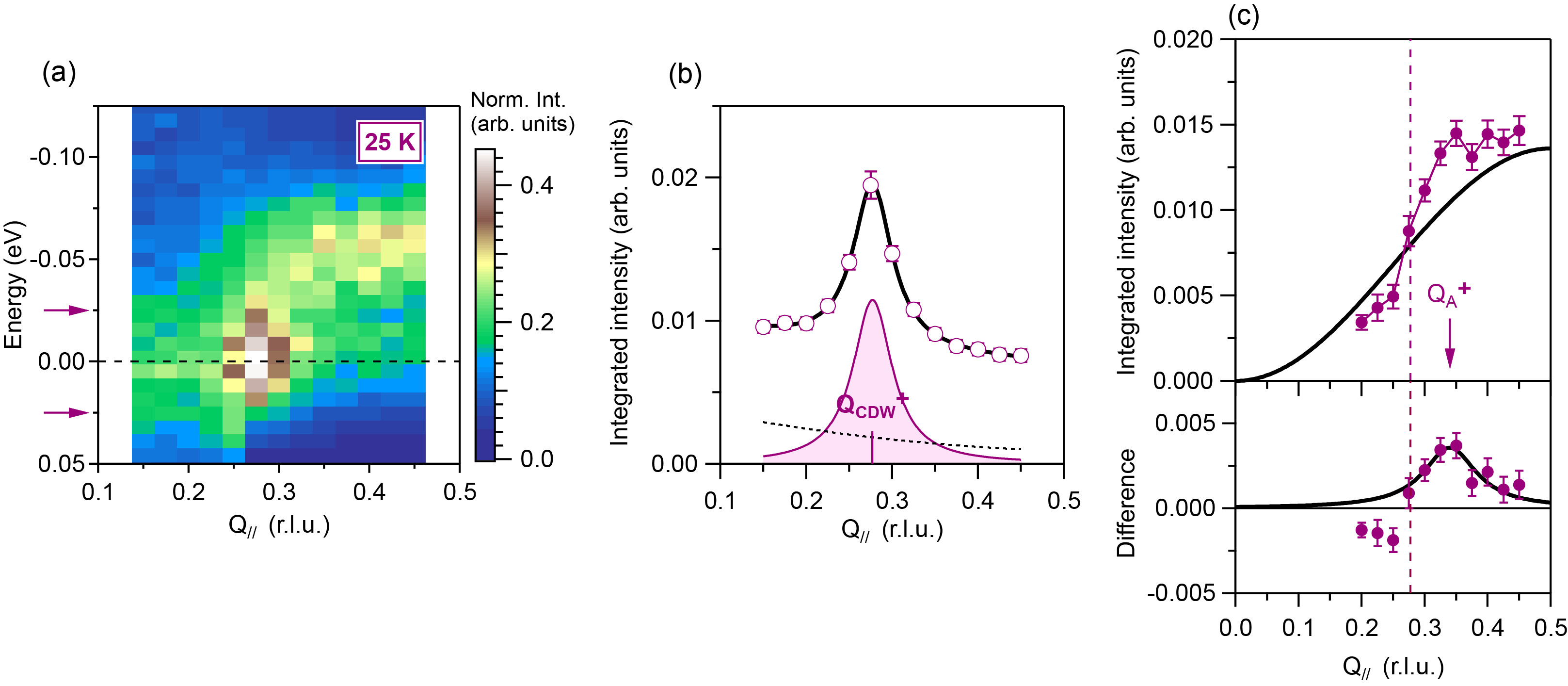}}
\caption{(a) RIXS intensity map recorded at 25 K with low statistics, as function of the energy and $Q_{\parallel}$. (b) Momentum-dependence of the integrated quasi-elastic intensity of the positive $Q_{\parallel}$ data at 25 K. The quasi-elastic region is defined as between the two purple arrows on (a). The black solid line is a fit to the data consisting of two Lorentzians, one for the CDW and the other for the tail of the specular reflection peak at $Q_{\parallel}$ = 0, plus a constant. The fit of the CDW peak is highlighted in purple. The CDW wave-vector is indicated with a vertical solid line. The background corresponding to the tail of the specular reflection peak is shown with a black dotted line. (c) Momentum-dependence of the RIXS integrated intensity of the elastic/paramagnon-subtracted positive $Q_{\parallel}$ data at 25 K. The integrated energy window is 0-100 meV. The black solid line is the $\text{sin}^{2}(\pi Q_{\parallel})$ curve. The bottom panel represents the difference between the integrated intensity and the $\text{sin}^{2}(\pi Q_{\parallel})$ curve, with Lorentzian fit (solid line). The dashed line indicates $Q_{\text{CDW}}^{\dag}$. The arrow highlights the anomaly wave-vector $Q_{\text{A}}^{\dag}$.}
\label{fig10}
\end{figure*}

RIXS data recorded at 25 K and positive $Q_{\parallel}$ in Na-CCOC sample 1 are presented in Fig. \ref{fig10}. Note that the energy resolution was approximately $\Delta$E $\sim$ 60 meV, except for the five first scans ($i.e.$ $Q_{\parallel}$ = 0.15 to $Q_{\parallel}$ = 0.25 r.l.u.), where the energy resolution was $\Delta$E $\sim$ 80 meV, due to a beam loss. As expected, measurements at positive $Q_{\parallel}$ confirm the observations of the CDW at $Q_{\text{CDW}}^{\dag} \sim 0.28 \pm 0.01$ r.l.u. [Fig. \ref{fig10} (b)], with a FWHM$^{\dag}$ $\sim 0.06 \pm 0.01$ r.l.u., corresponding to a correlation length of $\xi^{\dag}$ $\sim 20$ \AA. Note that the correlation lengths have been calculated from $\xi$ = $\frac{\text{a}}{2\pi\text{ HWHM}}$. 
Concerning the Cu-O bond stretching phonon, positive $Q_{\parallel}$ measurements confirm an intensity anomaly at $Q_{\text{A}}^{\dag}$ $\sim 0.34 \pm 0.02$ r.l.u. [Fig. \ref{fig10} (c)], despite a lower statistics than the data taken at negative $Q_{\parallel}$. Table \ref{tab1} summarizes the CDW and intensity anomaly wave-vectors, found from the analysis of the positive $Q_{\parallel}$ data. This comparison between positive and negative $Q_{\parallel}$ data enables solid crosschecking of the observations in two independent measurements.

\end{document}